\catcode`\@=11
\newif\if@fewtab\@fewtabtrue

{\count255=\time\divide\count255 by 60
\xdef\hourmin{\number\count255}
\multiply\count255 by-60\advance\count255 by\time
\xdef\hourmin{\hourmin:\ifnum\count255<10 0\fi\the\count255}}
\def\ps@draft{\let\@mkboth\@gobbletwo
    \def\@oddhead{}
    \def\@oddfoot
       {\hbox to 7 cm{$\scriptstyle Draft\ version:\ \draftdate$
       \hfil}\hskip -7cm\hfil\rm\thepage \hfil}
    \def\@evenhead{}\let\@evenfoot\@oddfoot}

\def\ceqno{\global\@fewtabfalse
    \ifcase\@eqcnt \def\@tempa{& & &}\or \def\@tempa{& &}
      \or \def\@tempa{&}
      \or\def\@tempa{}\fi\@tempa
{\rm(\theequation)}}

\def\aeqno#1{\global\@fewtabfalse
    \ifcase\@eqcnt \def\@tempa{& & &}\or \def\@tempa{& &}
      \or \def\@tempa{&}
      \or\def\@tempa{}\fi\@tempa
{\rm(\theequation,#1)}}

\def\label#1{\ifnum\draftcontrol=1
 \global\def\draftnote{$\scriptstyle #1$}\fi
 \@bsphack\if@filesw {\let\thepage\relax
   \def\protect{\noexpand\noexpand\noexpand}%
\xdef\@gtempa{\write\@auxout{\string
      \newlabel{#1}{{\@currentlabel}{\thepage}}}}}\@gtempa
   \if@nobreak \ifvmode\nobreak\fi\fi\fi
  \@esphack}

\def\alabel#1#2{\label{#1}\global\@fewtabfalse
    \ifcase\@eqcnt \def\@tempa{& & &}\or \def\@tempa{& &}
      \or \def\@tempa{&}
      \or\def\@tempa{}\fi\@tempa
{\hbox to 3cm{\phantom{\rm(\theequation,#2)}
\draftnote \hfil}\hskip -3cm {\rm(\theequation,#2)}}}

\def\clabel#1{\label{#1}\global\@fewtabfalse
    \ifcase\@eqcnt \def\@tempa{& & &}\or \def\@tempa{& &}
      \or \def\@tempa{&}
      \or\def\@tempa{}\fi\@tempa
{\hbox to 3cm{\phantom{\rm(\theequation)}
\draftnote \hfil}\hskip -3cm{\rm(\theequation)}}}

\def\eqnarray{\def\draftnote{{}}\global\@fewtabtrue
\stepcounter{equation}\let\@currentlabel=\theequation
\global\@eqnswtrue
\global\@eqcnt\z@\tabskip\@centering\let\\=\@eqncr
$$\halign to \displaywidth\bgroup\@eqnsel\hskip\@centering\@eqcnt\z@
  $\displaystyle\tabskip\z@{##}$&\global\@eqcnt\@ne
  \hskip 1\arraycolsep \hfil${##}$\hfil
  &\global\@eqcnt\tw@ \hskip 1\arraycolsep
$\displaystyle\tabskip\z@{##}$
\hfil  \tabskip\@centering&\global\@eqcnt\thr@@\llap{##}\tabskip\z@
\cr}

\def\endeqnarray{\@@eqncr\egroup
      \global\advance\c@equation\m@ne$$\global\@ignoretrue}

\def\@eqnnum{\hbox to 3cm{\phantom{\rm(\theequation)} \draftnote
                         \hfil}\hskip -3cm {\rm(\theequation)}}

\def\@@eqncr{\let\@tempa\relax
    \ifcase\@eqcnt \def\@tempa{& & &}\or \def\@tempa{& &}
      \or \def\@tempa{&}
      \or\def\@tempa{}
\fi\@tempa
\if@eqnsw
\if@fewtab\@eqnnum\fi
\stepcounter{equation}\fi\global
\@eqnswtrue\global\@eqcnt\z@\global\@fewtabtrue\cr}

\def\draftcite#1{\ifnum\draftcontrol=1#1\else{}\fi}

\def\@lbibitem[#1]#2{\item{}\hskip -3cm \hbox to 2cm
{\hfil$\scriptstyle\draftcite{#2}$}\hskip
1cm[\@biblabel{#1}]\if@filesw
     {\def\protect##1{\string ##1\space}\immediate
      \write\@auxout{\string\bibcite{#2}{#1}}}\fi\ignorespaces}

\def\@bibitem#1{\item\hskip -3cm \hbox to 2cm
{\hfil $\scriptstyle\draftcite{#1}$}\hskip 1cm
\if@filesw \immediate\write\@auxout
       {\string\bibcite{#1}{\the\value{\@listctr}}}\fi\ignorespaces}

\def\nsection#1{\section{#1}\setcounter{equation}{0}}

\font\tendl=msbm10  scaled \magstep1
\font\sevendl=msbm7 scaled \magstep1
\font\fivedl=msbm5 scaled \magstep1
\font\tengl=eufm10  scaled \magstep1
\font\sevengl=eufm7 scaled \magstep1
\font\fivegl=eufm5 scaled \magstep1

\newfam\dlfam  
\textfont\dlfam=\tendl \scriptfont\dlfam=\sevendl
\scriptscriptfont\dlfam=\fivedl
\newfam\glfam  
\textfont\glfam=\tengl \scriptfont\glfam=\sevengl
\scriptscriptfont\glfam=\fivegl

\def\draftdate{\number\month/\number\day/\number\year\ \ \ \hourmin }

\global\def\draftcontrol{0}
\catcode`\@=12

\def\hat{\widehat}

\documentstyle[11pt]{article}

\renewcommand{\theequation}{\thesection.\arabic{equation}}
\def\theequation{{\thesection.\arabic{equation}}}

\newcommand{\be}{\begin{eqnarray}}
\newcommand{\en}{\end{eqnarray}\vs 0.5 cm}

\newcommand{\vs}{\vskip}

\newcommand{\qq}{\begin{eqnarray}}

\newcommand{\qqq}{\end{eqnarray}}

\begin{document}
\title{\bf{SCALING\ LAWS\ IN\ TURBULENCE}\footnote{lecture 
given at the NATO Advanced Research Workshop ``New Developments in Field
Theory'', Zakopane, Poland, June 14-20, 1997}}
\author{\ \\Krzysztof Gaw\c{e}dzki \\ I.H.E.S., C.N.R.S.,
F-91440  Bures-sur-Yvette, France}
\date{ }
\maketitle

\vskip 0.3cm
\vskip 1 cm

\begin{abstract}
\vskip 0.3cm

\noindent The largely open problem of scaling laws in fully
developed turbulence is discussed, with the stress
put on similarities and differences with scaling in field theory. 
A soluble model of the passive advection is examined in more detail 
in order to illustrate the principal ideas. 
\end{abstract}
\vs 2cm

\nsection{LANGEVIN VERSUS NAVIER-STOKES}
\setcounter{equation}{0}
\vskip 0.5cm

Many dynamical problems in physics may be described 
by evolution equations of the type
\begin{eqnarray}
\partial_t\Phi{\hspace{0.05cm}}={\hspace{0.05cm}}-{\hspace{0.025cm}}
F(\Phi){\hspace{0.05cm}}
+{\hspace{0.05cm}} f
\label{ee}
\end{eqnarray}
where $\Phi(t,{{\bf x}})$ represents local densities
of physical quantities, $F(\Phi)$ is their nonlinear 
functional and $f$ stands for an external source. 
We shall be interested in the situations where the source
$f$ is random. For concreteness, we shall assume
it Gaussian with mean zero and covariance
\begin{eqnarray}
\langle{\hspace{0.05cm}} f(t,{{\bf x}}){\hspace{0.05cm}} f(s,{\bf y})
{\hspace{0.05cm}}\rangle{\hspace{0.05cm}}
={\hspace{0.05cm}}\delta(t-s){\hspace{0.05cm}}{\cal C}({_{{{\bf x}}
-{\bf y}}\over^L})
\label{cf}
\end{eqnarray}
with $L$ determining the scale on which $f(t,{{\bf x}})$
are correlated. An example of such an evolution law is provided 
by the Langevin equation describing the approach to equilibrium 
in systems of statistical mechanics or field theory {\cite{HH}}. 
In this case the nonlinearity is of the gradient type:
\begin{eqnarray}
F(\Phi){\hspace{0.05cm}}={\hspace{0.05cm}}{{\delta S(\Phi)}\over{\delta\Phi}}
\end{eqnarray}
with $S(\Phi)$ a local functional, 
e.g. $S(\Phi)={_1\over^2}\int(\nabla\Phi)^2+{_1\over^2} m^2\int\Phi^2
+\lambda\int\Phi^4$ in the $\Phi^4$ field theory,
and with $L$ small so that ${\cal C}({{\bf x}}/L)$
is close to the delta-function $\delta({{\bf x}})$ and regulates 
the theory on short distances 
$\mathop{<}\limits_{^\sim}{\hspace{0.025cm}}L$.
\vskip 0.3cm

Augmented by an initial condition $\Phi(t_0,{{\bf x}})$ 
(and, eventually, boundary conditions), the solution
of Eq.{\hspace{0.1cm}}(\ref{ee}) should define random field $\Phi(t,
{{\bf x}})
{\hspace{0.025cm}}$.
{\hspace{0.025cm}}We are interested in the behavior of its correlation 
functions given by the mean values
\begin{eqnarray}
\langle{\hspace{0.05cm}}\prod\limits_i\Phi(t_i,{{\bf x}}_i){\hspace{0.05cm}}
\rangle{\hspace{0.05cm}}.
\end{eqnarray}
Among the basic questions one may ask are the following ones:
\vskip 0.1cm

1. {\hspace{0.05cm}} Do the correlation 
functions become stationary (i.e.{\hspace{0.15cm}}dependent
only on time differences) when $t_0\to-\infty{\hspace{0.025cm}}$?
If so, are the stationary correlators unique (independent
of the initial condition)?
\vskip 0.1cm

2. {\hspace{0.05cm}}Do they obey scaling laws?
\vskip 0.2cm

\noindent For the field theory case these questions are well studied
with the use of powerful analytic tools as perturbative
expansions and renormalization group and by numerical
analysis (Monte Carlo simulations).
The stationary correlators describe possibly different
phases of the system. Universal (i.e.\hspace{0.15cm}independent of
the cutoff ${\cal C}{\hspace{0.025cm}}$) {\hspace{0.025cm}}scaling 
laws of the type
\begin{eqnarray}
\langle{\hspace{0.05cm}} Q(\Phi(t,{\bf x})){\hspace{0.05cm}}
 Q(\Phi(t,{\bf y})){\hspace{0.05cm}}\rangle\ \sim\ 
|{\bf x}-{\bf y}|^{-\zeta_Q}
\end{eqnarray}
for $|{\bf x}-{\bf y}|\gg L$ with $Q$ some local functions
of $\Phi$ emerge at the points of the $2^{
{\hspace{0.025cm}}{\rm nd}}$ 
order phase transitions. 
\vskip 0.3cm

On the opposite pole of the field theoretic case are the hydrodynamical
examples of the evolution equation (\ref{ee}). The best known
of those is the Navier-Stokes equation
\begin{eqnarray}
\partial_t {\bf v}{\hspace{0.05cm}}={\hspace{0.05cm}}
-P({\bf v}\cdot\nabla){\bf v}{\hspace{0.05cm}}
+{\hspace{0.05cm}}\nu\Delta {\bf v}
{\hspace{0.05cm}}+{\hspace{0.05cm}}{\bf f}
\label{NS}
\end{eqnarray}
for the incompressible ($\nabla\cdot{\bf v}=0$) velocity field 
${\bf v}(t,{\bf x})
{\hspace{0.025cm}}$,
{\hspace{0.025cm}}with $P$ standing for the orthogonal 
projection on such vector fields.
$\nu$ denotes the viscosity and ${\bf f}$ is the external
force which induces the fluid motion. In the fully developed turbulence 
one is interested in the regime where the stirring forces act
on large distances (like the convective forces on scales
of kilometers in the atmosphere) and we observe quite complicated
(turbulent) motions on shorter distances down to scales 
on which the dissipative term $\propto\nu$ becomes important 
({\hspace{0.025cm}}$\sim$ milimeters in the atmosphere). 
It is believed that the large scale details
should not be essential for the statistics of the flow
in this intermediate regime called the ``inertial range''. It is therefore
common to model the stirring forces by a random Gaussian process 
with mean zero and covariance
\begin{eqnarray}
\langle{\hspace{0.05cm}} f^\alpha(t,{\bf x}){\hspace{0.05cm}}
 f^\beta(s,{\bf y}){\hspace{0.05cm}}\rangle{\hspace{0.05cm}}
={\hspace{0.05cm}}\delta(t-s){\hspace{0.05cm}}
{\cal C}^{\alpha\beta}({_{{\bf x}-{\bf y}}\over^L})
\end{eqnarray}
with $\partial_\alpha{\cal C}^{\alpha\beta}=0{\hspace{0.025cm}}$. $L$ 
denotes now the large "integral scale" on which the random forces act.
Note that, unlike for field theory, in this case the covariance 
${\cal C}({\bf x}/L)$ is close to a constant, i.e.{\hspace{0.15cm}}to 
a delta-function in the wavenumber space and not 
in the position space. Such regime in field theory would 
correspond to distances shorter than the ultraviolet cutoff
with the behavior strongly dependent on the detailed form 
of the regularization. Another (related) difference is that 
in Eq.{\hspace{0.1cm}}(\ref{NS}) the nonlinear term is not of a gradient 
type. Finally, the projection $P$ renders it non-local which 
is another complication. All these differences make 
the Navier-Stokes problem (\ref{NS}) quite different from that 
posed by the Langevin equation and resistant to the methods
employed successfully in the study of the latter.

\nsection{KOLMOGOROV THEORY}

The first major attempt to obtain universal scaling laws
for the inertial range correlators is due to 
Kolmogorov {\cite{K41}}. Assuming the existence of homogeneous 
(i.e.{\hspace{0.15cm}}translationally 
invariant) stationary correlators
of velocities, one deduces the following relation at equal times
\begin{eqnarray} 
{_1\over^4}\nabla_{\bf x}\cdot\langle{\hspace{0.05cm}}
({\bf v}({\bf x})-{\bf v}({\bf y}))^2
{\hspace{0.05cm}}
({\bf v}({\bf x})-{\bf v}({\bf y})){\hspace{0.05cm}}\rangle{\hspace{0.05cm}}
+{\hspace{0.05cm}}{_1\over^2}{\hspace{0.05cm}}{\rm tr}
{\hspace{0.1cm}}{\cal C}({_{{\bf x}-{\bf y}}
\over^{L}}){\hspace{0.05cm}}={\hspace{0.05cm}}
\nu{\hspace{0.05cm}}\langle{\hspace{0.05cm}}
\nabla{\bf v}({\bf x})\cdot\nabla{\bf v}({\bf y})
{\hspace{0.05cm}}\rangle{\hspace{0.05cm}}.
\label{k1}
\end{eqnarray}
Eq.{\hspace{0.1cm}}(\ref{k1}) is obtained the following
way. First, using Eq.{\hspace{0.1cm}}(\ref{NS}), we write
\begin{eqnarray}
{\bf v}(t+\delta t,{\bf x}){\hspace{0.05cm}}
={\bf v}(t,{\bf x}){\hspace{0.05cm}}+
{\hspace{0.05cm}}[-P({\bf v}\cdot
\nabla){\bf v}(t,{\bf x}){\hspace{0.05cm}}+
{\hspace{0.05cm}}
\nu\Delta{\bf v}(t,{\bf x})]\delta t{\hspace{0.05cm}}+
\int\limits_t^{t+
\delta t}{\bf f}(s){\hspace{0.05cm}} ds
{\hspace{0.05cm}}+{\hspace{0.05cm}}{\cal O}(\delta t^2)
\hspace{0.3cm}
\end{eqnarray}
and we instert this expansion into the equal-time 2-point
function obtaining
\begin{eqnarray}
&&\langle{\hspace{0.05cm}}{\bf v}(t+\delta t,{\bf x})
{\hspace{0.05cm}}{\bf v}(t+\delta t,{\bf y})
{\hspace{0.05cm}}\rangle{\hspace{0.05cm}}={\hspace{0.05cm}}
\langle{\hspace{0.05cm}}{\bf v}(t,{\bf x}){\hspace{0.05cm}}
{\bf v}(t,{\bf y}){\hspace{0.05cm}}\rangle\cr
&&-{\hspace{0.05cm}}\langle{\hspace{0.05cm}}
 P({\bf v}\cdot\nabla){\bf v}(t,{\bf x})\cdot{\bf v}(t,{\bf y})
{\hspace{0.05cm}}
\rangle\delta t-{\hspace{0.05cm}}\langle{\hspace{0.05cm}}
{\bf v}(t,{\bf x})\cdot 
P({\bf v}\cdot\nabla)
{\bf v}(t,{\bf y}{\hspace{0.05cm}}\rangle\delta t \cr
&&+{\hspace{0.05cm}}\nu\langle{\hspace{0.05cm}}
\Delta{\bf v}(t,{\bf x})\cdot{\bf v}(t,{\bf y}){\hspace{0.05cm}}\rangle
\delta t{\hspace{0.05cm}}+{\hspace{0.05cm}}\nu\langle
{\hspace{0.05cm}}{\bf v}(t,{\bf x})\cdot\Delta{\bf v}(t,{\bf y})
{\hspace{0.05cm}}
\rangle\delta t{\hspace{0.05cm}}+{\hspace{0.05cm}}{\rm tr}{\hspace{0.1cm}}
{\cal C}({_{{\bf x}-{\bf y}}
\over^{L}})\delta t{\hspace{0.05cm}}+{\hspace{0.05cm}}{\cal O}(\delta t^2)
{\hspace{0.05cm}}.
\end{eqnarray}
Vanishing of the ${\cal O}(\delta t)$ terms 
produces Eq.{\hspace{0.1cm}}(\ref{k1}).
\vskip 0.35cm

Under the limit ${\bf y}\to{\bf x}{\hspace{0.025cm}}$ for positive $\nu$, 
Eq.\hspace{0.1cm}(\ref{k1}) becomes the identity
\begin{eqnarray}
{_1\over^2}{\hspace{0.05cm}}{\rm tr}{\hspace{0.1cm}}{\cal C}
(0)\ =\ \langle{\hspace{0.05cm}}
\nu(\nabla{\bf v})^2{\hspace{0.025cm}}
\rangle{\hspace{0.05cm}}\equiv{\hspace{0.05cm}}\bar\epsilon
\end{eqnarray}
which expresses the energy balance: in the stationary state,
the mean energy injection rate is equal to the mean rate of
energy dissipation $\bar\epsilon{\hspace{0.025cm}}$. 
On the other hand, performing
the limit $\nu\to0$ for ${\bf x}\not={\bf y}$ one obtains
\begin{eqnarray}
-{\hspace{0.025cm}}{_1\over^4}\nabla_{\bf x}
\cdot\langle{\hspace{0.05cm}}({\bf v}({\bf x})-{\bf v}({\bf y}))^2
{\hspace{0.05cm}}
({\bf v}({\bf x})-{\bf v}({\bf y})){\hspace{0.05cm}}\rangle\ 
=\ {_1\over^2}{\hspace{0.05cm}}{\rm tr}
{\hspace{0.1cm}}{\cal C}({_{{\bf x}-{\bf y}}\over^L})
{\hspace{0.05cm}}.
\label{cc}
\end{eqnarray}
For $\vert{\bf x}-{\bf y}\vert\ll L$ the right hand side 
is approximately constant and equal to ${_1\over^2}{\hspace{0.05cm}}{\rm tr}
{\hspace{0.1cm}}{\cal C}(0){\hspace{0.025cm}}$, 
{\hspace{0.025cm}}i.e.{\hspace{0.15cm}}to $\bar\epsilon{\hspace{0.025cm}}$. 
Assuming also isotropy (rotational invariance) of the stationary
state one may then infer the form of the 3-point function 
in the inertial range:
\begin{eqnarray}
\langle{\hspace{0.05cm}}(v^\alpha({\bf x})-v^\alpha(0))
{\hspace{0.05cm}}(v^\beta({\bf x})-v^\beta(0))
{\hspace{0.05cm}}(v^\gamma({\bf x})-v^\gamma(0)){\hspace{0.05cm}}\rangle
\ \cong\ -{4{\hspace{0.025cm}}\bar\epsilon\over d(d+2)}
{\hspace{0.05cm}}(\delta^{\alpha\beta} x^\gamma{\hspace{0.05cm}}
+{\hspace{0.05cm}}{\rm cycl.}){\hspace{0.05cm}}.\hspace{0.5cm}
\end{eqnarray}
The latter implies for the 3-point function of the component
of $({\bf v}({\bf x})-{\bf v}(0))$ paralel to ${\bf x}$ the relation
\begin{eqnarray}
\langle{\hspace{0.05cm}}({\bf v}(x)
-{\bf v}(0))_{_{\vert\vert}}^{{\hspace{0.05cm}}3}
{\hspace{0.025cm}}
\rangle\ \cong\ -{\hspace{0.025cm}}{12\over d(d+2)}
{\hspace{0.075cm}}{\hspace{0.025cm}}\bar
\epsilon{\hspace{0.075cm}}\vert{\bf x}\vert
\label{4/5}
\end{eqnarray}
known as the Kolmogorov "$4\over 5$ law"
($4\over 5$ is the value of the coeffitient on the right hand 
side in 3 dimensions).
\vskip 0.3cm

Under natural assumptions about the stationary state one may
deduce a stronger version (sometimes called refined
similarity) of the above relation which takes
the form of the operator product expansion
for the $\nu\to0$ limit of the dissipation operator 
$\epsilon=\nu{\hspace{0.025cm}}(\nabla{\bf v})^2{\hspace{0.025cm}}$:
\begin{eqnarray}
\epsilon(x)\ =\ -{\hspace{0.025cm}}{_1\over^4}{\hspace{0.075cm}}
\lim\limits_{{\bf y}\to{\bf x}}
{\hspace{0.05cm}}
{\hspace{0.05cm}}\nabla_x\cdot[({\bf v}({\bf x})
-{\bf v}({\bf y}))^2{\hspace{0.05cm}}
({\bf v}({\bf x})-{\bf v}({\bf y}))]{\hspace{0.05cm}}\bigg\vert_{{\nu=0}}
\label{da}
\end{eqnarray}
holding inside the expectations in the $\nu\to0$ limit. 
Relation (\ref{da}) expresses the dissipative anomaly:
the dissipation $\epsilon$ whose definition involves
a factor of $\nu$ does not vanish when $\nu\to0{\hspace{0.025cm}}$.
\vskip 0.3cm

Kolmogorov postulated {\cite{K41}} that 
the scaling of general velocity 
correlators in the inertial range should be determined 
by universal relations involving only the distances and 
the mean dissipation rate $\bar\epsilon{\hspace{0.025cm}}$. Such postulate 
leads to the scaling laws
\begin{eqnarray}
\langle{\hspace{0.05cm}}({\bf v}({\bf x})
-{\bf v}(0))_{_{\vert\vert}}^{{\hspace{0.05cm}}
n}{\hspace{0.025cm}}\rangle
\ {\hspace{0.05cm}}\propto{\hspace{0.05cm}}\ \bar\epsilon^{n/3}
{\hspace{0.05cm}}\vert{\bf x}\vert^{n/3}
\label{kn}
\end{eqnarray}
for the n-point "structure functions" of velocity generalizing 
the (essentially rigorous) result (\ref{4/5}) about
the 3-point function. The right hand side of  relation (\ref{kn})
is the only expression built from $\bar\epsilon$ and 
$\vert{\bf x}\vert$ with the right dimension. 
\vskip 0.3cm

The power law fits $\propto{\hspace{0.025cm}}\vert{\bf x}\vert^{\zeta_n}$
for the structure functions measured in experiments and 
in numerical simulations lead to the values of the
exponents slightly different from the Kolmogorov prediction
$\zeta_n=n/3$ for $n\not=3{\hspace{0.025cm}}$. {\hspace{0.025cm}}One 
obtains {\cite{Benzi}}
$\zeta_2\cong .70{\hspace{0.025cm}}$, $\zeta_4\cong 1.28
{\hspace{0.025cm}}$, 
$\zeta_6\cong 1.77{\hspace{0.025cm}}$, $\zeta_8\cong 2.23
{\hspace{0.025cm}}$.{\hspace{0.05cm}} 
The discrepencies indicate that the random variables
$({\bf v}({\bf x})-{\bf v}(0))_{_{\vert\vert}}{\hspace{0.025cm}}$ 
are non-Gaussian
for small ${\bf x}$ with the probability distribution
functions decaying slower than in the normal
distribution. Such a slow decay signals the phenomenon of frequent occurence 
of large deviations from the mean values called "intermittency".
There exist many phenomenological models of intermittency 
of the inertial range velocity differences 
based on the idea that the turbulent activity is carried 
by a fraction of degrees of freedom with a self-similar 
("multi-fractal") structure {\cite{Frisch}}. An explanation 
of the mechanism behind the observed intermittency 
starting from the first principles (i.e.\hspace{0.15cm}from the Navier-Stokes
equation) is, however, still missing and constitutes the main 
open fundamental problem of the fully developed turbulence.

\nsection{KRAICHNAN MODEL OF PASSIVE ADVECTION}

Recently some progress has been 
achieved {\cite{KG},\cite{Falk1},\cite{ShrS}} 
in understanding the origin of intermittency in a simple 
model {\cite{Kr68},\cite{Kr94}}
describing advection of a scalar quantity (temperature 
{\hspace{0.025cm}}$T(t,{\bf x}){\hspace{0.025cm}}$)
{\hspace{0.025cm}}by a random velocity field 
${\bf v}(t,{\bf x}){\hspace{0.025cm}}$. 
{\hspace{0.025cm}}The evolution of the temperature is described 
by the equation
\begin{eqnarray}
\partial_t T{\hspace{0.05cm}}={\hspace{0.05cm}} -
{\hspace{0.025cm}}({\bf v}\cdot\nabla)T{\hspace{0.05cm}}
+{\hspace{0.05cm}}\kappa\Delta T{\hspace{0.05cm}}+{\hspace{0.05cm}} f
\label{ps}
\end{eqnarray}
where $\kappa$ denotes the molecular diffusivity and $f$
is the external source which we shall take random Gaussian with
mean zero and covariance (\ref{cf}). Following 
Kraichnan {\cite{Kr68}},
we shall assume that ${\bf v}(t,{\bf x})$ is also a Gaussian process, 
independent of $f{\hspace{0.025cm}}$, 
{\hspace{0.025cm}}with mean zero and covariance 
\begin{eqnarray}
\langle{\hspace{0.05cm}} v^\alpha(t,{\bf x}){\hspace{0.05cm}}
 v^\beta(s,{\bf y}){\hspace{0.05cm}}\rangle\ =\ 
\delta(t-s){\hspace{0.1cm}}(D_0\delta^{\alpha\beta}-
{\hspace{0.025cm}} d^{\alpha\beta}
({\bf x}-{\bf y}){\hspace{0.025cm}})
\end{eqnarray}
with $D_0$ a constant, $d^{\alpha\beta}({\bf x}){\hspace{0.025cm}}
\propto{\hspace{0.025cm}}\vert{\bf x}\vert^\xi$ for small $\vert{\bf x}\vert$
and with $\partial_\alpha d^{\alpha\beta}=0$ in order to assure
the incompressibility. Note the scaling of the 2n-point function
of velocity differences with power $n\xi$ of the distance.
The Komogorov scaling corresponds to $\xi={4\over 3}$
(the temporal delta-function appears to have dimension 
$length^{\xi-1}$). The time decorrelation 
of the velocities is not, however, a very realistic assumption.
In the Kraichnan model $\xi$ is treated as a parameter
running from $0$ to $2{\hspace{0.025cm}}$. 
\vskip 0.3cm

Writing
\begin{eqnarray}
T(t+\delta t,{\bf x}){\hspace{0.025cm}}={\hspace{0.025cm}} T(t,{\bf x})
-\hspace{-0.1cm}\int\limits_t^{t+\delta t}\hspace{-0.1cm}
[{\hspace{0.025cm}}
({\bf v}\cdot\nabla)T(s,{\bf x}){\hspace{0.025cm}}
-{\hspace{0.025cm}} f(s,{\bf x}){\hspace{0.025cm}}]{\hspace{0.05cm}} ds
{\hspace{0.025cm}}+
{\hspace{0.025cm}}\kappa\Delta T(t,{\bf x})\delta t
{\hspace{0.025cm}}+{\hspace{0.025cm}}{\cal O}(\delta t^2)
{\hspace{0.025cm}},\hspace{0.4cm}
\label{exp}
\end{eqnarray}
we obtain the analogue of the relation (\ref{k1}) for 
the stationary state of the scalar (the latter may be shown to exist
and to be independent of the initial condition decaying at spatial
infinity):
\begin{eqnarray}
-{\hspace{0.025cm}}{_1\over^2}{\hspace{0.025cm}}
 d^{\alpha\beta}({\bf x}-{\bf y}){\hspace{0.1cm}}
\partial_{x^\alpha}\partial_{y^\beta}
{\hspace{0.05cm}}\langle{\hspace{0.05cm}}
T({\bf x}){\hspace{0.05cm}} T({\bf y}){\hspace{0.05cm}}
\rangle{\hspace{0.05cm}}+{\hspace{0.05cm}}{_1\over^2}
{\hspace{0.025cm}}{\cal C}({_{{\bf x}-{\bf y}}
\over^L}){\hspace{0.05cm}}={\hspace{0.05cm}}\kappa{\hspace{0.05cm}}
\langle{\hspace{0.05cm}}\nabla T({\bf x})\cdot\nabla T({\bf y})
{\hspace{0.05cm}}\rangle{\hspace{0.05cm}}.
\label{kr1}
\end{eqnarray}
Letting in Eq.{\hspace{0.05cm}}(\ref{kr1}) ${\bf y}\to{\bf x}$ for $\kappa>0$
produces, as before, the energy balance:
\begin{eqnarray}
{_1\over^2}{\hspace{0.025cm}}
{\cal C}(0){\hspace{0.05cm}}={\hspace{0.05cm}}\langle{\hspace{0.05cm}}
\kappa{\hspace{0.025cm}}(\nabla T)^2{\hspace{0.025cm}}\rangle{\hspace{0.05cm}}
\equiv{\hspace{0.05cm}}\bar\epsilon{\hspace{0.05cm}}.
\end{eqnarray}
On the other hand, taking $\kappa\to 0$ for ${\bf x}\not={\bf y}$
results in the equation analogous to (\ref{cc}):
\begin{eqnarray}
{_1\over^2}{\hspace{0.025cm}} 
d^{\alpha\beta}({\bf x}-{\bf y}){\hspace{0.1cm}}
\partial_{x^\alpha}
\partial_{y^\beta}{\hspace{0.05cm}}\langle{\hspace{0.05cm}} T({\bf x})
{\hspace{0.05cm}} T({\bf y}){\hspace{0.05cm}}\rangle{\hspace{0.05cm}}
={\hspace{0.05cm}}{_1\over^2}{\hspace{0.025cm}}{\cal C}
({_{{\bf x}-{\bf y}}\over^L})
\end{eqnarray}
which may be easily solved exactly for the 2-point function 
of the scalar giving
\begin{eqnarray}
\langle{\hspace{0.05cm}}
 T({\bf x}){\hspace{0.05cm}} T(0){\hspace{0.025cm}}
\rangle{\hspace{0.05cm}}={\hspace{0.05cm}}
A_{\cal C}{\hspace{0.05cm}} L^{2-\xi}{\hspace{0.05cm}}-{\hspace{0.05cm}}
{\rm const}.{\hspace{0.1cm}}\bar\epsilon{\hspace{0.05cm}}
{\hspace{0.025cm}}\vert{\bf x}\vert^{2-\xi}
{\hspace{0.05cm}}
+{\hspace{0.05cm}}{\cal O}(L^{-2})
\label{kr2}
\end{eqnarray}
or
\begin{eqnarray}
\langle{\hspace{0.05cm}}(T({\bf x})-T(0))^2
{\hspace{0.025cm}}\rangle{\hspace{0.05cm}}
\cong{\hspace{0.05cm}}{\rm const}.{\hspace{0.1cm}}
\bar\epsilon
{\hspace{0.075cm}}\vert{\bf x}\vert^{2-\xi}
\label{kr3}
\end{eqnarray}
for $\vert{\bf x}\vert\ll L{\hspace{0.025cm}}$. {\hspace{0.05cm}}This is 
an analogue of the
Kolmogorov {\hspace{0.025cm}}${4\over 5}{\hspace{0.025cm}}$ law. 
It may be strengthen
to the operator product expansion for the dissipation
operator {\cite{Proc1},\cite{BKG}}
$\epsilon=\kappa(\nabla T)^2$
\begin{eqnarray}
\epsilon(x)\ =\ {_1\over^2}\hspace{0.15cm}
\lim\limits_{{\bf y}\to{\bf x}}{\hspace{0.1cm}}
d^{\alpha\beta}({\bf x}-{\bf y}){\hspace{0.1cm}}
\partial_{\alpha}T({\bf x}){\hspace{0.075cm}}
\partial_{\beta}T({\bf y}){\hspace{0.05cm}}\bigg\vert_{{\kappa=0}}
\label{das}
\end{eqnarray}
valid inside the expectations in the limit $\kappa\to0{\hspace{0.025cm}}$.
Eq.{\hspace{0.1cm}}(\ref{das}) expresses the dissipative anomaly
in the Kraichnan model, analogous to the dissipative anomaly
(\ref{da}) for the Navier-Stokes case.
\vskip 0.3cm

The natural question arises whether the higher structure
functions of the scalar $\langle{\hspace{0.05cm}}(T({\bf x})-T(0))^{2n}
{\hspace{0.025cm}}\rangle
\equiv S_{2n}({\bf x})$ scale with powers $n(2-\xi)$
as the dimensional analysis would suggest (Corrsin's 
analogue {\cite{Corrs}} of the Kolmogorov theory). 
The answer is no. Experiments show that the scalar
differences display higher intermittency than that
of the velocities {\cite{AHGA}}.
Although, by assumption, in the Kraichnan model there is 
no intermittency in the distribution of the velocity
differences, numerical 
studies {\cite{KYC},\cite{FGLP}} indicate strong 
intermittency of the scalar differences signaled by anomalous 
(i.e.\hspace{0.15cm}$\not=n(2-\xi))$ scaling 
exponents. Unlike in the Navier-Stokes case, we have now some analytic 
understanding of this phenomenon, although still incomplete and 
controvertial {\cite{Kr94},\cite{FGLP},\cite{Kr97}}.
\vskip 0.3cm

The simplifying feature of the Kraichnan model is that
the insertion of expansion (\ref{exp}) into the higher point 
functions ${\cal F}_{2n}(\underline{\bf x})\equiv\langle
{\hspace{0.025cm}} T(t,{\bf x}_1,\dots, 
T(t,{\bf x}_{2n})\rangle$ leads to a system of (Hopf) equations 
which close:
\begin{eqnarray}
{\cal M}_{2n}{\hspace{0.05cm}}{\cal F}_{2n}
(\underline{{\bf x}})\ =\ \sum\limits_{i<j}{\cal F}_{2n-2}
(\mathop{{\bf x}_1,\dots\dots,{\bf x}_{2n}}\limits_{\hat{i}\ \ \ \hat{j}})
{\hspace{0.1cm}}{\cal C}({_{{\bf x}_i-{\bf x}_j}\over^L})
\label{Hopf}
\end{eqnarray}
where ${\cal M}_{n}$ are differential operators
\begin{eqnarray}
{\cal M}_{n}=\sum\limits_{1\leq i<j\leq n}d^{\alpha\beta}({\bf x}_i-{\bf x}_j)
{\hspace{0.1cm}}
\partial_{x_i^\alpha}{\hspace{0.05cm}}\partial_{x_j^\beta}{\hspace{0.05cm}}
-{\hspace{0.05cm}}\kappa
\sum\limits_{i=1}^n\Delta_{{\bf x}_i}{\hspace{0.05cm}}.
\label{Mn}
\end{eqnarray}
In principle, the above equations permit to determine uniquely 
the stationary higher-point correlators
of the scalar iteratively by inverting the positive elliptic 
operators ${\cal M}_{2n}{\hspace{0.025cm}}$. By analyzing these operators
whose symbols loose strict positivity when $\kappa$ goes
to zero, it was argued in {\cite{KG},\cite{BKG}} 
that at least for small $\xi$
\vskip 0.1cm

1. \ $\lim\limits_{\kappa\to0}{\hspace{0.1cm}}
{\cal F}_{2n}(\underline{{\bf x}})$
exists and is finite,    
\vskip 0.1cm

2. \ ${\cal F}_{2n}(\underline{{\bf x}})\ =\ A_{{\cal C},2n}{\hspace{0.05cm}}
 L^{\rho_{2n}}{\hspace{0.075cm}}{\cal F}^0_{2n}
(\underline{{\bf x}}){\hspace{0.05cm}}+{\hspace{0.05cm}}
{\cal O}(L^{-2+{\cal O}(\xi)}){\hspace{0.05cm}}+{\hspace{0.05cm}}\dots$\ \ 
\ \ at $\kappa=0$ 
\vskip 0.2cm
\noindent where $\rho_{2n}={2n(n-1)\over d+2}{\hspace{0.025cm}}\xi
{\hspace{0.05cm}}+{\hspace{0.05cm}}{\cal O}(\xi^2){\hspace{0.025cm}}$,
${\cal F}^0_{2n}$ is a scaling zero mode of the $\kappa=0$
version ${\cal M}_{2n}^0$ of the operator ${\cal M}_{2n}{\hspace{0.025cm}}$: 
${\cal M}_{2n}^0{\hspace{0.05cm}}{\cal F}^0_{2n}=0{\hspace{0.025cm}}$,
${\cal F}^0_{2n}(\lambda\underline{{\bf x}})=\lambda^{n(2-\xi)-\rho_{2n}}
{\hspace{0.075cm}}{\cal F}^0_{2n}(\underline{{\bf x}}){\hspace{0.025cm}}$. 
{\hspace{0.025cm}}The 
{\hspace{0.025cm}}$\dots$-terms do not depend
on at least one of the vectors ${\bf x}_i$ and do not
contribute to the correlators of scalar differences.
\vskip 0.1cm

3. \ $S_{2n}({\bf x})\ \propto\ L^{\rho_{2n}}{\hspace{0.075cm}}\vert{\bf x}
\vert^{n(2-\xi)-\rho_{2n}}$ \ \ at $\kappa=0$ and
for $\vert{\bf x}\vert\ll L{\hspace{0.025cm}}$.
\vskip 0.1cm

\noindent The last relation, a simple consequence of the second one, 
shows appearence of intermittent exponents at least for small 
$\xi$ ({\hspace{0.025cm}}$\rho_{2n}$ is their anomalous part and 
it is positive starting from the 4-point function).
A similar analysis, consistent with the above one, has been 
done for large space dimensions {\cite{Falk1},\cite{Falk2}}.
\vskip 0.3cm

The above results about the "zero-mode dominance" 
of the correlators of the scalar differences show
what degree of universality one may expect in the 
scaling laws of intermittent quantities: the amplitudes
$A_{{\cal C},2n}$ in front of the dominant term depend
on the shape of the covariance ${\cal C}$ i.e.{\hspace{0.15cm}}on 
the details of the large scale stirring. But the zero 
modes of the dominant terms (and their scaling exponents) 
do not. In field theory, small-scale universality
of the critical behavior finds its explication in the
renormalization group analysis. Similarly, in 
the Kraichnan model there exists a renormalization
group explanation of the observed long scale
universality {\cite{MG}}. The renormalization group
transformations eliminate subsequently the long scale 
degrees of freedom. In a sense, they consist of looking
at the system by stronger and stronger magnifying glass
so that the long distance details are lost from sight. 
The eliminated degrees of freedom induce an effective
source for the remaining ones {\cite{MG}}. Whereas such 
an "inverse renormalization group" analysis may be implemented 
for more complicated turbulent systems rests an open problem.

\nsection{DYNAMICS OF LAGRANGIAN TRAJECTORIES}

What is the source of the zero mode dominance of the
inertial range correlators of the scalar differences?
In absence of the diffusion term in Eq.{\hspace{0.1cm}}(\ref{ps})
the scalar density is given by the integral
\begin{eqnarray}
T(t,{\bf x}){\hspace{0.05cm}}={\hspace{0.05cm}}\int\limits_{-\infty}^t 
f(s;{\bf y}(s;t,{\bf x})){\hspace{0.075cm}} ds
\end{eqnarray}
where ${\bf y}(s;t,{\bf x})$ describes the Lagrangian trajectory,
i.e.\hspace{0.15cm}the solution of the equation
\begin{eqnarray}
{d{\bf y}\over ds}{\hspace{0.05cm}}
={\hspace{0.05cm}}{\bf v}(s,{\bf y}){\hspace{0.05cm}},
\label{lt}
\end{eqnarray}
passing at time $t$ through point ${\bf x}$
(for concretness, we have assumed the vanishing initial condition
for $T$ at $t=-\infty{\hspace{0.025cm}}$). 
{\hspace{0.025cm}}The Lagrangian trajectories
describe the flow of the fluid elements. If the velocities
are random, so are Lagrangian trajectories and we
may ask the question about their joint probability
distributions. Let $P_{n}(t,\underline{{\bf x}};
{\hspace{0.025cm}} s,\underline{{\bf y}})$
denote the probability that $n$ Lagrangian trajectories
starting at time $s$ at points $({\bf y}_1,\dots,{\bf y}_n)\equiv
\underline{{\bf y}}$ pass at time $t\geq s$ through points 
$({\bf x}_1,\dots,{\bf x}_n)\equiv\underline{{\bf x}}$. 
These probabilities may be
computed for the time-decorrelated Gaussian velocities.
They appear to be given by the heat kernels of the singular 
elliptic operators ${\cal M}_n^0{\hspace{0.025cm}}$:
\begin{eqnarray}
P_n(t,\underline{{\bf x}};{\hspace{0.025cm}} s,\underline{{\bf y}})\ =\ 
{\rm e}^{-(t-s){\cal M}_n^0}(\underline{{\bf x}};\underline{{\bf y}})
\end{eqnarray}
(more exactly, this is true after averaging over the simultaneous
translations of the initial points, i.e.\hspace{0.15cm}for the probabilities
of relative positions of the trajectories). 
\vskip 0.3cm

From the form of the operators ${\cal M}_n^0$ we see that the
(relative positions of) $n$ Lagrangian trajectories undergo 
a diffusion process with distance-dependent diffusion coefficients.
When two trajectories are close, the corresponding diffusion
coefficient vanishes as the distance to power $\xi{\hspace{0.025cm}}$
slowing down the diffusive separation of the trajectories.
When the trajectories eventually separate, the diffusion
coefficient grows speeding up further separation. 
The result is a superdiffusive large time asymptotics:
\begin{eqnarray}
\int f(\underline{{\bf x}}){\hspace{0.1cm}} 
P_n(t,\underline{{\bf x}};{\hspace{0.025cm}}0,\underline{{\bf y}})
{\hspace{0.1cm}} d\underline{{\bf x}}
\ \ \propto\ \ t^{\sigma\over2-\xi}
\label{sd}
\end{eqnarray}
for a generic (translationally invariant) scaling function
$f(\underline{{\bf x}})$ of scaling dimension $\sigma>0$
(e.g.{\hspace{0.1cm}}for $\sum_{i,j}|{\bf x}_i-{\bf x}_j|^2$ 
with the scaling dimension $2{\hspace{0.025cm}}$). Note the faster 
than diffusive growth in time for 
$\xi>0{\hspace{0.025cm}}$. There are, however,
exceptions from this generic behavior. In particular, if
$f$ is a scaling zero mode of ${\cal M}_n^0$ then
\begin{eqnarray}
\int f(\underline{{\bf x}}){\hspace{0.1cm}} P_n(t,\underline{{\bf x}};
{\hspace{0.025cm}} 0,\underline{{\bf y}})
{\hspace{0.1cm}} d\underline{{\bf x}}
\ =\ \int f(\underline{{\bf x}}){\hspace{0.1cm}}{\rm e}^{-t
{\hspace{0.025cm}}{\cal M}_n^0}(\underline{{\bf x}};\underline{{\bf y}})
{\hspace{0.1cm}} d\underline{{\bf x}}
\ ={\rm const}.\ =\ f(\underline{{\bf y}}){\hspace{0.05cm}}.
\label{sd1}
\end{eqnarray}
It can be shown {\cite{SM}} that each 
zero mode $f_0$ of ${\cal M}_n^0$
of scaling dimension $\sigma_0 \geq 0$ 
generates descendent slow collective modes 
$f_p{\hspace{0.025cm}}$, {\hspace{0.025cm}} 
p=1,2,\dots, \, of scaling dimensions
$\sigma_p=\sigma_0+p(2-\xi)$ for which 
\begin{eqnarray}
\int f_p(\underline{{\bf x}}){\hspace{0.1cm}} 
P_n(t,\underline{{\bf x}};{\hspace{0.025cm}} 0,\underline{{\bf y}})
{\hspace{0.1cm}} d\underline{{\bf x}}
\ \ \propto\ \ t^p{\hspace{0.05cm}},
\label{sd2}
\end{eqnarray}
i.e.{\hspace{0.15cm}}grows slower (if $\sigma_0>0{\hspace{0.025cm}}$) 
{\hspace{0.025cm}}than in (\ref{sd}).
The descendants satisfy the chain of equations ${\cal M}_n f_p
=f_{p-1}{\hspace{0.025cm}}$. {\hspace{0.025cm}}The 
structure with towers of descendants
over the primary zero modes resembles that in systems
with infinite symmetries and may suggest presence of
hidden symmetries in the Kraichnan model.
\vskip 0.3cm

The slow modes $f_p$ appear in the asymptotic expansion
\begin{eqnarray}
P_n(t,\underline{{\bf x}};{\hspace{0.025cm}} 0,\underline{{\bf y}}/L)\ 
=\ \sum\limits_{{\rm zero\ modes}\atop f_0}
\sum\limits_{p=0}^\infty L^{-\sigma_p}
{\hspace{0.075cm}} \overline{g_p(t,\underline{{\bf x}})}
{\hspace{0.075cm}} f_p(\underline{{\bf y}})
\label{ase}
\end{eqnarray}
valid for large $L{\hspace{0.025cm}}$ and describing the behavior of the
trajectories starting close to each other. Since 
$P_n(t,\underline{{\bf x}};{\hspace{0.025cm}} 0,\underline{{\bf y}}/L)
=P_n(t,\underline{{\bf y}}/L;{\hspace{0.025cm}} 0,
\underline{{\bf x}}){\hspace{0.025cm}}$,
expansion (\ref{ase}) describes also probabilities that the
trajectories approach each other after time $t{\hspace{0.025cm}}$.
By simple rescalings and the use of expansion (\ref{ase}), 
one obtains
\begin{eqnarray}
\int f(\underline{{\bf x}}){\hspace{0.1cm}}
P_n(t,\underline{{\bf x}};{\hspace{0.025cm}} 0,\underline{{\bf y}})
{\hspace{0.1cm}} d\underline{{\bf x}}
{\hspace{0.05cm}}={\hspace{0.05cm}} t^{\sigma\over 2-\xi}
\int f({\bf x}){\hspace{0.1cm}} P_n(1,\underline{{\bf x}};
{\hspace{0.025cm}} 0,\underline{{\bf y}}/t^{1\over 2-\xi}){\hspace{0.1cm}} 
d\underline{{\bf x}}\cr
={\hspace{0.05cm}}\sum\limits_{{\rm zero\ modes}\atop f_0}
\sum\limits_{p=0}^\infty t^{{\sigma-\sigma_p\over 2-\xi}}{\hspace{0.075cm}}
f_p(\underline{{\bf y}})\int f(\underline{{\bf x}}){\hspace{0.075cm}} 
\overline{g_p(t,\underline{{\bf x}})}{\hspace{0.075cm}} 
d\underline{{\bf x}}{\hspace{0.05cm}}.
\end{eqnarray}
For generic $f{\hspace{0.025cm}}$, {\hspace{0.025cm}}the dominant 
term comes from the
constant zero mode and is proportional to
$t^{\sigma\over 2-\xi}{\hspace{0.025cm}}$. 
{\hspace{0.025cm}}However for $f$ equal to one of 
the slow modes, the leading contribution is given by the 
further terms in the expansion due to the orthogonality relations
between $f_p{\hspace{0.025cm}}$'s and $g_p{\hspace{0.025cm}}$'s. 
The behaviors 
(\ref{sd},\ref{sd1},\ref{sd2}) result. The distributions 
$P_n(t,\underline{{\bf x}};{\hspace{0.025cm}} 0,
\underline{{\bf y}})$ enter the expressions for
scalar correlators. For those of scalar differences, 
the contributions to expansion (\ref{ase}) of modes which do not
depend on all ${\bf y}_i$ (like the constant mode) drop out
leading to the dominance by non-trivial zero modes.
\vskip 0.3cm

The asymptotic expansion (\ref{ase}), governed by the slow modes
$f_p$, displays another important feature of the Lagrangian
trajectories. Since $P_n(t,\underline{{\bf x}};
{\hspace{0.025cm}} 0,\underline{{\bf y}})$
is a joint probability distribution of the endpoints of
$n$ Lagrangian trajectories starting at points 
${\bf y}_1,\dots,{\bf y}_n{\hspace{0.025cm}}$, we should expect that 
it becomes concentrated on the diagonal
${\bf x}_1=\cdots={\bf x}_n$ when the initial points ${\bf y}_1,\dots,
{\bf y}_n$ tend to each other. But this is not the case as
$\lim\limits_{L\to\infty}{\hspace{0.05cm}}
P_n(t,\underline{{\bf x}};{\hspace{0.025cm}} 0,
\underline{{\bf y}}/L)=P_n(t,\underline{{\bf x}};
{\hspace{0.025cm}} 0,0)$
is a finite function of 
$\underline{{\bf x}}{\hspace{0.025cm}}$. 
\hspace{0.025cm}It is the constant zero mode contribution to expansion
(\ref{ase}). The whole expansion describes how exactly 
$P_n(t,\underline{{\bf x}};{\hspace{0.025cm}} 0,
\underline{{\bf y}}/L)$ fails to concentrate
on the diagonal when $L\to\infty{\hspace{0.025cm}}$. 
{\hspace{0.025cm}}Similarly, $P_n(t,0;
{\hspace{0.025cm}} 0,\underline{{\bf y}})$ is a non-singular
function of $\underline{{\bf y}}$ showing that the probability
that the trajectories collapse after time $t$ to
a single point is finite, contradicting the uniqueness of
the Lagrangian trajectories passing through a given point.
The solution of the paradox is as follows. The typical 
velocities in the ensemble that we consider have rough spatial 
behavior (and even rougher time behavior). As functions 
of ${\bf x}$ they are (essentially) H\"older continuous 
with exponent $\xi\over 2{\hspace{0.025cm}}$. 
{\hspace{0.025cm}}But for such velocities, 
the equation for the Lagrangian trajectories (\ref{lt}) does 
not have a unique solution, given the initial position.
As a result, the Lagrangian trajectories loose deterministic 
character for a fixed velocity realization. Nevertheless, one may 
still talk about probability distribution 
$P_n(t,\underline{{\bf x}};{\hspace{0.025cm}} 0,
\underline{{\bf y}}{\hspace{0.025cm}}
|{\hspace{0.025cm}}{\bf v})$ of their final
points whose average over ${\bf v}$ gives 
$P_n(t,\underline{{\bf x}};{\hspace{0.025cm}} 0,
\underline{{\bf y}}){\hspace{0.025cm}}$.
{\hspace{0.025cm}}In a more realistic description which takes into
account a smoothing of the typical velocities
at very short viscous scale, the same effect is due
to the sensitive dependence of the now deterministic
Lagrangian trajectories on the initial conditions,
within the viscous scale, signaled by the positivity of the 
Lyapunov exponent {\cite{SM}}.
\vskip 0.3cm

Summarizing, intermittency in the Kraichnan model
of the passive advection appears to be due to the 
slow collective modes in the otherwise superdiffusive 
stochastic Lagrangian flow. The presence of such modes is 
closely related to the breakdown of the deterministic 
character of Lagrangian trajectories for the fixed 
velocity configuration at high Reynolds numbers, due 
to the sensitive dependence of the trajectories 
on the initial conditions within the viscous scale. 
We expect both phenomena to be present also in more realistic 
turbulent velocity distributions and to be still responsible 
for the anomalous scaling and intermittency.

\end{document}